\documentclass[a4paper, conference]{IEEEtran}
\IEEEoverridecommandlockouts
\usepackage[utf8]{inputenc}
\usepackage{graphicx}
\usepackage{tikz}
\usepackage{pgfplots}
\usepackage{psfrag}
\usetikzlibrary{arrows,snakes,shapes,shapes.geometric,fit,positioning,arrows,decorations.markings,calc}
\usepackage{mathtools}
\usepackage{amsbsy}
\usepackage{amsmath}
\usepackage{fixmath}
\usepackage{amssymb}
\usepackage{mathrsfs}
\usepackage{fancyhdr}
\usepackage{xcolor}
\usepackage{hyperref}
\usepackage{algorithm}
\usepackage{algcompatible}
\usepackage{algpseudocode}
\usepackage{multirow}
\newcommand{\bs}[1]{\boldsymbol{#1}}

\newcommand{\HH}{\mathrm{H}}
\newcommand{\TT}{\mathrm{T}}

\usepackage{accents}

\newcommand{\tdl}{\text{dl}}
\DeclareMathOperator*{\argmin}{arg\,min}

\definecolor{lime}{HTML}{A6CE39}
\DeclareRobustCommand{\orcidicon}{%
	\begin{tikzpicture}
	\draw[lime, fill=lime] (0,0) 
	circle [radius=0.16] 
	node[white] {{\fontfamily{qag}\selectfont \tiny ID}};
	\draw[white, fill=white] (-0.0625,0.095) 
	circle [radius=0.007];
	\end{tikzpicture}
	\hspace{-2mm}
}

\foreach \x in {A, ..., Z}{%
	\expandafter\xdef\csname orcid\x\endcsname{\noexpand\href{https://orcid.org/\csname orcidauthor\x\endcsname}{\noexpand\orcidicon}}
}

\newcommand{\copyrighttext}{%
\footnotesize\textcopyright This work has been submitted to the IEEE for possible publication. Copyright may be transferred without notice, after which this version may no longer be accessible.}

\newcommand{\copyrightnotice}{%
\begin{tikzpicture}[remember picture,overlay]
\node[anchor=south,yshift=10pt] at (current page.south) {\fbox{\parbox[]{\dimexpr\textwidth-\fboxsep-\fboxrule\relax}{\copyrighttext}}};
\end{tikzpicture}%
}

\makeatletter
\def\endthebibliography{%
	\def\@noitemerr{\@latex@warning{Empty `thebibliography' environment}}%
	\endlist
}
\makeatother

\begin{document}
	
	\title{Asymptotic Behavior of Zero-Forcing Precoding based on Imperfect Channel Knowledge for Massive MISO FDD Systems 
	}
	
	\author{\IEEEauthorblockN{ Donia~Ben~Amor\orcidA{}, Michael~Joham\orcidB{}, Wolfgang~Utschick\orcidC{} }
		\IEEEauthorblockA{ 	\textit{Chair of Methods of Signal Processing} \\
			\textit{School of Computation, Information and Technology, Technical University of Munich} \\
			Munich, Germany \\
			Email: \{donia.ben-amor, joham, utschick\}@tum.de}
	}
	\maketitle

\copyrightnotice
	\begin{abstract}
		In this work, we study the asymptotic behavior of the zero-forcing precoder based on the least squares (LS) and the linear minimum mean-square error (LMMSE) channel estimates for the downlink (DL) of a frequency-division-duplex (FDD) massive multiple-input-single-output (MISO) system.
		We show analytically the rather surprising result that zero-forcing precoding based on the LS estimate leads asymptotically to an interference-free transmission, even if the number of pilots used for DL channel training is less than the number of antennas available at the base station (BS). Although the LMMSE channel estimate exhibits a better quality in terms of the MSE due to the exploitation of the channel statistics, we show that in the case of contaminated channel observations, zero-forcing based on the LMMSE is unable to eliminate the inter-user interference in the asymptotic limit of high DL transmit powers. In order for the results to hold, mild conditions on the channel probing phase are assumed. The validity of our analytical results is demonstrated through numerical simulations for different scenarios.
	\end{abstract}
	
	
	\section{Introduction}
	Massive multiple-input-multiple-output (MaMIMO) also known as large-scale antenna \cite{marzetta}, \cite{marzetta2016fundamentals} is expected to be an essential component of future cellular networks due to its various advantages, which include increased spectral and energy efficiency \cite{mimo}. These benefits are, however, strongly dependent on the transmit strategies deployed at the BS in the case of a DL setup. Linear precoding schemes are often used in order to reduce the computational complexity and are shown to achieve relatively good performance compared to the more complex non-linear techniques \cite{Joham}, \cite{Stojnic}. One suboptimal, yet simple, commonly used precoding scheme is the zero-forcing precoder which was shown to be asymptotically optimal in the high SNR regime \cite{CaireShamai}, \cite{Wiesel}. This asymptotic behavior is, however, only guaranteed in case of perfect channel state information (CSI) at the transmitter. When it comes to an FDD operated MaMIMO system, the acquisition of CSI is a challenging task due to the short channel coherence interval on the one hand. On the other hand, the channel reciprocity usually assumed in time-division-duplex (TDD) systems does not hold \cite{Tse}. This means that the BS has to send known pilots to the users, which in turn have to estimate their own channels and feed them back to the BS in the uplink (UL). In order to reduce the training overhead particularly for MaMIMO systems, we assume in this work that the number of DL pilots is less than the number of BS antennas (the latter is the minimum number required for a contamination-free estimate). We show analytically and numerically that even in the case of contaminated channel observations, the zero-forcing precoder based on the LS channel estimate leads to an interference-free transmission at high transmit powers. Furthermore, we prove that the better quality estimator, namely the LMMSE estimator, does not exhibit the same behavior and fails to mitigate the inter-user interference even in the high power regime. In other words, if certain conditions on the DL channel probing phase are satisfied, the zero-forcing based on the less demanding LS channel estimate asymptotically provides full degrees of freedom. Hence, the channel statistics needed for the computation of the LMMSE channel estimate are not required to be known and one can, therefore, save computational resources. This can be beneficial especially for an FDD setup, where, for example, the DL channel covariance matrices have be additionally estimated either from an additional training phase or by performing an extrapolation approach from the UL channel covariance matrices \cite{Donia_WSA20}, \cite{Extrapolation2}. Based on these results, one should be aware of the fact that a good quality channel estimate does not necessarily yield a satisfactory overall system performance and that simpler estimation techniques can possibly deliver a higher system throughput. In this work, we concentrate on the more general and realistic case where the channels are spatially correlated. In the case the channels are uncorrelated (this corresponds to a scaled identity as channel covariance matrix), the LS and LMMSE channel estimates coincide and no further insights can be gained from this specific case.
	\par This paper is organized as follows: In Section~\ref{sec:2}, we present the system model for the considered setup and review the LS and LMMSE channel estimation together with the zero-forcing precoder. The asymptotic behavior of the latter based on both channel estimates is studied analytically in Section~\ref{sec:3}. These results are validated through numerical results in Section~\ref{sec:4}.
	\section{System Model and Setup}\label{sec:2}
	We consider the DL of a single-cell massive MISO system operating in FDD mode. The BS is equipped with many antennas $M\gg 1$ and serves $K$ single-antenna users. The channel between the BS and user $k$ is denoted by $\bs{h}_k\sim\mathcal{N}_\mathbb{C}(\bs{0},\bs{C}_k)$, with $\bs{C}_k$ being the channel covariance matrix of user $k$.
	\par Due to the non-reciprocity of the UL and DL channels in the FDD setup, the BS has to send pilots to the users which have to feed back their channel estimates to the BS via a feedback link. In order to reduce the training overhead, we consider the case where the number of pilots $T_\tdl$ used for channel probing is much smaller than the minimum number required for a contamination-free estimation (in the DL case the latter coincides with the number of BS antennas $M$). The $T_\tdl$ channel observations at user $k$ can be collected in the vector $\bs{y}_k$
	\begin{equation} \label{eq:yk}
		\bs{y}_k=\bs{\Phi}^\HH \bs{h}_k + \bs{n}_k
	\end{equation}
	where $\bs{\Phi} \in \mathbb{C}^{M\times T_\text{dl}}$ denotes the pilot matrix, whose columns are given by the $T_\text{dl}$ orthogonal pilot sequences sent by the BS to the users for the sake of channel probing. Due to the orthogonality of the pilot sequences, it holds that $\bs{\Phi}^\HH\bs{\Phi}=\bs{I}_{T_\text{dl}}$, but $\bs{\Phi}\bs{\Phi}^\HH \neq \bs{I}_{M}$, since $M>T_\tdl$.\\  
	For simplicity, we assume that the users feed back their observations, i.e., $\bs{y}_k, k=1,\dots, K$ without feedback errors. In this case, the noise vector $\bs{n}_k \sim \mathcal{N}_\mathbb{C}(\bs{0},\sigma^2\bs{I}_{T_\tdl})$ captures the DL training noise whose variance is given by $\sigma^2=1/P_\text{dl}$, where $P_\text{dl}$ denotes the DL transmit power.
	\subsection{Channel Estimation}
	Based on the channel observation vector $\bs{y}_k$, the BS can compute the least squares (LS) channel estimate $\hat{\bs{h}}_k^\text{LS}$ of user $k$, which is given by
	\begin{equation}
		\hat{\bs{h}}_k^\text{LS}= \underset{\hat{\bs{h}}_k}{\argmin}\: \|\bs{y}_k-\bs{\Phi}^\HH 	\hat{\bs{h}}_k\|^2=(\bs{\Phi}\bs{\Phi}^\HH)^\dagger\bs{\Phi} \bs{y}_k
	\end{equation}
	where $\bs{A}^\dagger$ denotes the pseudo-inverse of a matrix $\bs{A}$.
	Due to the property $\bs{\Phi}^\HH\bs{\Phi}=\bs{I}_{T_\text{dl}}$, it holds that $(\bs{\Phi}\bs{\Phi}^\HH)^\dagger=\bs{\Phi}\bs{\Phi}^\HH$. Thus, the LS estimate can be rewritten as
	\begin{equation}\label{eq:LS}
		\hat{\bs{h}}_k^\text{LS}= \bs{\Phi}\bs{\Phi}^\HH \bs{\Phi} \bs{y}_k =  \bs{\Phi} \bs{y}_k.
	\end{equation}
	
	In addition to the LS estimate, the BS can compute the linear minimum mean squared error (LMMSE) channel estimate based on the training observation $\bs{y}_k$ and the channel covariance matrix $\bs{C}_k$. The LMMSE estimate can be written as 	$\hat{\bs{h}}_k^\text{LMMSE}= \bs{G}_k^\ast \bs{y}_k$, where the matrix $\bs{G}_k^\ast$ is given by
	\begin{align*}
		\bs{G}_k^\ast &=\underset{\bs{G}_k}{\argmin} \: \mathbb{E}[ \| \bs{h}_k - \bs{G}_k \bs{y}_k\|^2]\\
		&= \bs{C}_k \bs{\Phi} \left(\bs{\Phi}^\HH \bs{C}_k \bs{\Phi} +\sigma^2 \bs{I}_{T_\tdl}\right)^{-1}.
	\end{align*}
	The LMMSE estimate of the $k$th user's channel hence reads as
	\begin{equation}\label{eq:LMMSE}
		\hat{\bs{h}}_k^\text{LMMSE}=\bs{C}_k \bs{\Phi} \left(\bs{\Phi}^\HH \bs{C}_k \bs{\Phi} +\sigma^2 \bs{I}_{T_\tdl}\right)^{-1} \bs{y}_k.
	\end{equation}
	\subsection{Data Transmission and Zero-Forcing Precoding}
	During the data transmission phase, the BS sends precoded signals to the $K$ users. The transmit precoded vector is given by
	\begin{equation}
		\bs{x}=\sum_{k=1}^{K} \bs{p}_k s_k = \bs{P} \bs{s} 
	\end{equation}
	with the precoding matrix $\bs{P}=[\bs{p}_1, \dots, \bs{p}_K]$ and the precoder $\bs{p}_k$ for user $k$. $ s_k \sim \mathcal{N}_\mathbb{C}(0,1)$ is the encoded data signal intended for user $k$ and $\bs{s}=[s_1, \dots, s_K]^\TT$. 
	
	Hence, the signal received at user $k$ reads as
	\begin{equation}
		r_k= \bs{h}_k^\HH \bs{p}_k  s_k +  \sum_{j\neq k} \bs{h}_k^\HH \bs{p}_j s_j+v_k
	\end{equation}
	with the normalized additive white Gaussian noise signal $v_k\sim \mathcal{N}_\mathbb{C}(0,\sigma^2)$.
	
	In this work, we focus on the case where zero-forcing precoding is employed. Given the estimated channel matrix $\hat{\bs{H}}=[\hat{\bs{h}}_1, \dots, \hat{\bs{h}}_K]$, the zero-forcing precoding matrix is computed according to
	\begin{equation}\label{eq:ZFPrec}
		\bs{P}^\text{ZF}=\delta (\hat{\bs{H}}^\HH)^\dagger = \delta \hat{\bs{H}} (\hat{\bs{H}}^\HH \hat{\bs{H}})^{-1}. 
	\end{equation}
	Here the scaling factor $\delta=\frac{1}{\sqrt{\text{tr}((\hat{\bs{H}}^\HH \hat{\bs{H}})^{-1})}}$ is introduced to normalize the precoding matrix, such that $\text{tr}(\bs{P}^\text{ZF} \bs{P}^{\text{ZF},\HH})=1$.\\
	Note that the matrix $\hat{\bs{H}}^\HH \hat{\bs{H}}$  should be full-rank in order for the zero-forcing precoding matrix to exist. This is in general the case, if $K\leq M$. Specifically for the LS estimate, we can write the estimated channel matrix as
	\begin{equation}
		\hat{\bs{H}}=\bs{\Phi}\bs{Y}, \quad \text{with} \: \bs{Y}=[\bs{y}_1, \dots, \bs{y}_K] \label{eq:HhatLS}
	\end{equation}
	Thus, the matrix $\hat{\bs{H}}^\HH \hat{\bs{H}}$ can be written as
	\begin{equation}
		\hat{\bs{H}}^\HH \hat{\bs{H}}=\bs{Y}^\HH \bs{\Phi}^\HH \bs{\Phi}\bs{Y}=\bs{Y}^\HH\bs{Y}. \label{eq:HHHe}
	\end{equation}
	The zero-forcing precoding matrix based on the LS estimate exists therefore if $\bs{Y}^\HH\bs{Y}$ is invertible, which is the case if $K\leq T_\tdl$.
	
	In the next section, we will show that zero-forcing precoding based on the LS estimate leads to an interference-free transmission in the high power regime. Contrary, the interference in the high power regime cannot be fully suppressed when using the LMMSE estimate for zero-forcing precoding. For the following analysis, we assume that $M\geq T_\tdl$ which is in alignment with the considered massive MIMO setup, where the DL training overhead needs to be kept as low as possible.
	
	\section{Asymptotic Behavior of the Zero-Forcing Precoder}\label{sec:3}
	\subsection{Two-User Case}
	To simplify the analysis, mainly for the case when the LMMSE channel estimates are used, we concentrate on the 2-user case in this section, while the results are generalizable for $K>2$ (A proof for $K>2$ when zero-forcing precoding based on LS channel estimates is  presented in the next subsection). For the 2-user case, we have 
	\begin{equation} \label{eq:Hhat}
		\hat{\bs{H}}=[\hat{\bs{h}}_1,\hat{\bs{h}}_2]
	\end{equation}
	with the estimate $\hat{\bs{h}}_i$ for the $i$th channel.
	
	We study the asymptotic behavior of the system in the high power regime, i.e., $P_\tdl\rightarrow\infty$. To understand the interference behavior, we focus on the combination of one user's channel $\bs{h}_k$ with another user's precoding vector $\bs{p}_j, j\neq k$ given by $\bs{h}_k^\HH \bs{p}_j$. Remember that the received signal at user $k$ for the 2-user case reads as
	\begin{align}
		r_k= \bs{h}_k^\HH \bs{p}_k  s_k +  \bs{h}_k^\HH \bs{p}_j s_j+v_k.
	\end{align}
	We say that the interference is suppressed, when $\bs{h}_k^\HH \bs{p}_j=0$, otherwise, the system is interference-limited. 
	
	Before we proceed with our analysis, we rewrite the precoding matrix $\bs{P}^\text{ZF}$ as a function of the channel estimates $\hat{\bs{h}}_1$ and $\hat{\bs{h}}_2$ as follows [cf. \eqref{eq:ZFPrec} and \eqref{eq:Hhat}]
	\begin{align}
		\bs{P}^\text{ZF}&=[\bs{p}_1, \bs{p}_2] =\beta	[ \hat{\bs{h}}_1, \hat{\bs{h}}_2] \begin{bmatrix} \|\hat{\bs{h}}_2\|^2 & -\hat{\bs{h}}_1^\HH \hat{\bs{h}}_2 \\ -\hat{\bs{h}}_2^\HH \hat{\bs{h}}_1 & \|\hat{\bs{h}}_1\|^2 \end{bmatrix} \nonumber \\
		&=  \beta \begin{bmatrix}  \hat{\bs{h}}_1 \|\hat{\bs{h}}_2\|^2 -  \hat{\bs{h}}_2\hat{\bs{h}}_2^\HH \hat{\bs{h}}_1  & \hat{\bs{h}}_2 \|\hat{\bs{h}}_1\|^2 -  \hat{\bs{h}}_1\hat{\bs{h}}_1^\HH \hat{\bs{h}}_2  \end{bmatrix} \label{eq:p1p2}
	\end{align}
	with $\beta=\delta / \det( \hat{\bs{H}}^\HH  \hat{\bs{H}})=\delta/(\|\hat{\bs{h}}_1\|^2\|\hat{\bs{h}}_2\|^2 - |\hat{\bs{h}}_1^\HH \hat{\bs{h}}_2 |^2)$.
	
	\subsubsection{Asymptotic behavior with zero-forcing precoding based on LS estimate}
	
	We start with analyzing the case where the LS estimate is used for zero-forcing precoding.  Note that this analysis is only valid for the case when $T_\tdl \geq K$, because otherwise the zero-forcing precoder is not defined due to the non-existence of the inverse in \eqref{eq:ZFPrec}.\\
	First of all, since the training noise is inversely proportional to the DL transmit power (recall that $\sigma^2=\frac{1}{P_\tdl}$), the ``asymptotic'' LS  estimate, denoted by $	\hat{\bar{\bs{h}}}_k$, can be written as [see \eqref{eq:LS} and \eqref{eq:yk}]
	\begin{equation}
		\label{eq:LSasymp}
		\hat{\bar{\bs{h}}}_k=\bs{\Phi} \bs{\Phi}^\HH \bs{h}_k \quad \text{for} \: P_\tdl \rightarrow\infty.
	\end{equation}
	This can be seen by writing the noise vector $\bs{n}_k$ in the observation vector $\bs{y}_k$ as $\bs{n}_k=\sigma \tilde{\bs{n}}_k= \frac{1}{\sqrt{P_\tdl}}\tilde{\bs{n}}_k$, with $\tilde{\bs{n}}_k\sim \mathcal{N}_\mathbb{C}(\bs{0},\bs{I}_{T_\tdl})$. For $P_\tdl\rightarrow\infty$, the noise term can therefore be neglected.\\
	Let us now consider the interference caused by user 1 on user 2 in the asymptotic limit. To this end, we compute $\bs{h}_2^\HH \bs{p}_1$ for infinite $P_\tdl$ using the result from \eqref{eq:p1p2}.\\
	As for the channel estimate $\hat{\bs{h}}_k$, we deploy the asymptotic LS estimate $	\hat{\bar{\bs{h}}}_k$ in \eqref{eq:LSasymp} leading to
	\begin{align*}
		&\bs{h}_2^\HH \bs{p}_1=   \beta \bs{h}_2^\HH(	\hat{\bar{\bs{h}}}_1\|	\hat{\bar{\bs{h}}}_2\|2 -  	\hat{\bar{\bs{h}}}_2 	\hat{\bar{\bs{h}}}_2^\HH 	\hat{\bar{\bs{h}}}_1) \\
		&=  \beta   \bs{h}_2^\HH( \bs{\Phi} \bs{\Phi}^\HH \bs{h}_1 \bs{h}_2^\HH  \bs{\Phi} \underbrace{ \bs{\Phi}^\HH \bs{\Phi}}_{\bs{I}_{T_\tdl}} \bs{\Phi}^\HH \bs{h}_2 - \bs{\Phi} \bs{\Phi}^\HH \bs{h}_2 \bs{h}_2^\HH  \bs{\Phi} \underbrace{\bs{\Phi}^\HH \bs{\Phi} }_{\bs{I}_{T_\tdl}}\bs{\Phi}^\HH \bs{h}_1  )\\
		&= \beta  ( \bs{h}_2^\HH \bs{\Phi} \bs{\Phi}^\HH \bs{h}_1  \bs{h}_2^\HH  \bs{\Phi} \bs{\Phi}^\HH \bs{h}_2 -  \bs{h}_2^\HH 
			\bs{\Phi} \bs{\Phi}^\HH \bs{h}_2  \bs{h}_2^\HH  \bs{\Phi} \bs{\Phi}^\HH \bs{h}_1 )\\
		&=0.
	\end{align*} 
	Similarly, we can prove that $\bs{h}_1^\HH \bs{p}_2=0$ for $P_\tdl \rightarrow\infty$.\\
	Based on these results, we have shown that zero-forcing precoding based on the LS estimates leads asymptotically to an interference-free transmission.
	
	\subsubsection{Asymptotic behavior with zero-forcing precoding based on LMMSE estimate}
	Now, we study the asymptotic behavior of the system, when the LMMSE estimates are used to compute the zero-forcing precoding matrix. \\
	Note that for $P_\tdl\rightarrow\infty$ the LMMSE channel estimate can be written as [cf.~\eqref{eq:LMMSE} and \eqref{eq:yk} for $\sigma^2\rightarrow 0$]
	\begin{equation}
		\label{eq:LMMSEasymp}
		\hat{\bar{\bs{h}}}_k=\bs{C}_k \underbrace{\bs{\Phi} \left(\bs{\Phi}^\HH \bs{C}_k \bs{\Phi} \right)^{-1} \bs{\Phi}^\HH}_{\bs{C}_{\Phi k}} \bs{h}_k =\bs{C}_k \bs{C}_{\Phi k}\bs{h}_k.
	\end{equation}
	We again consider the term responsible for the interference gain $\bs{h}_2^\HH \bs{p}_1$ as $P_\tdl\rightarrow\infty$ using the channel estimates $	\hat{\bar{\bs{h}}}_k$ in \eqref{eq:LMMSEasymp} and the precoder expression \eqref{eq:p1p2}
	\begin{equation}
		\begin{aligned}
			\bs{h}_2^\HH \bs{p}_1 =& \beta \bs{h}_2^\HH( 	\hat{\bar{\bs{h}}}_1 \|	\hat{\bar{\bs{h}}}_2\|^2 -  	\hat{\bar{\bs{h}}}_2 	\hat{\bar{\bs{h}}}_2^\HH \	\hat{\bar{\bs{h}}}_1)\nonumber\\
			=& \beta 	\bs{h}_2^\HH (\bs{C}_1 \bs{C}_{\Phi 1}\bs{h}_1  \bs{h}_2^\HH \bs{C}_{\Phi 2}\bs{C}_2 \bs{C}_2 \bs{C}_{\Phi 2}\bs{h}_2\nonumber\\  &-  \bs{C}_2 \bs{C}_{\Phi 2}\bs{h}_2  \bs{h}_2^\HH \bs{C}_{\Phi 2}\bs{C}_2  \bs{C}_1 \bs{C}_{\Phi 1}\bs{h}_1 )\nonumber\\
			=& \beta  \bs{h}_2^\HH \bs{C}_{\Phi 2}\bs{C}_2 ( \bs{C}_2 \bs{C}_{\Phi 2}\bs{h}_2\bs{h}_2^\HH \nonumber \\
			&-\bs{h}_2^\HH \bs{C}_2 \bs{C}_{\Phi 2}\bs{h}_2 \bs{I}_M)\bs{C}_1 \bs{C}_{\Phi 1}\bs{h}_1\nonumber\\
			=&\beta  \bs{h}_2^\HH \bs{C}_{\Phi 2}\bs{C}_2 (\bs{C}_2 \bs{C}_{\Phi 2}\bs{h}_2\bs{h}_2^\HH \nonumber \\
			&-\text{tr}( \bs{C}_2 \bs{C}_{\Phi 2}\bs{h}_2 \bs{h}_2^\HH) \bs{I}_M)\bs{C}_1 \bs{C}_{\Phi 1}\bs{h}_1\\
			\neq & 0.
		\end{aligned} 
	\end{equation}
	Obviously, the term $\bs{h}_2^\HH \bs{p}_1$ does not converge to 0 when $P_\tdl\rightarrow\infty$. Therefore, there exists a residual interference when zero-forcing precoding based on LMMSE channel estimation is used.
	\subsection{Generalization of the Asymptotic Behavior for $K>2$}
	Using the compact formulation of the LS channel estimate in \eqref{eq:HhatLS}, we can show for the general case $K>2$, that the zero-forcing precoder based on the LS estimate leads to an interference-free transmission in the high power limit. Recall that the precoder is given by [cf. \eqref{eq:ZFPrec}]
	\begin{equation}
		\bs{P}^\text{ZF}=\delta \hat{\bs{H}}^\dagger = \delta \hat{\bs{H}} (\hat{\bs{H}}^\HH \hat{\bs{H}})^{-1}. \label{eq:ZFprec1}
	\end{equation}
	By inserting the LS estimate \eqref{eq:HhatLS} and\eqref{eq:HHHe} into \eqref{eq:ZFprec1}, we obtain
	\begin{equation*}
		\bs{P}^{\text{ZF}}=\delta \bs{\Phi} \bs{Y}(\bs{Y}^\HH \bs{Y})^{-1}.
	\end{equation*}
	Since the DL training noise variance is inversely proportional to the DL transmit power, the matrix $\bs{Y}$ of all channel observations [cf. \eqref{eq:HhatLS}] converges to $\bs{\Phi}^\HH\bs{H}$ for $P_\tdl\rightarrow\infty$. Thus, it holds for the zero-forcing matrix for $P_\tdl\rightarrow\infty$ that
	\begin{equation}
		\bs{P}^{\text{ZF,asy}}=\delta \bs{\Phi}\bs{\Phi}^\HH\bs{H} (\bs{H}^\HH \bs{\Phi}\bs{\Phi}^\HH\bs{H})^{-1}.
	\end{equation}
	Now,  we consider the effective channel given by the combination of the channel matrix with the precoding matrix, i.e.,
	\begin{equation} \label{eq:HhPzf}
		\bs{H}^\HH \bs{P}^{\text{ZF,asy}}=\delta \bs{H}^\HH \bs{\Phi}\bs{\Phi}^\HH\bs{H} (\bs{H}^\HH \bs{\Phi}\bs{\Phi}^\HH\bs{H})^{-1}=\delta \bs{I}_K.
	\end{equation}
	It is clear from \eqref{eq:HhPzf}, that the effective channel asymptotically  converges to a scaled identity, which provides a generalization of the behavior discussed in the previous subsection for $K=2$. For $P_\tdl\rightarrow\infty$, the zero-forcing precoder based on the LS estimate ensures an interference-free transmission if $T_\tdl\geq K$.
	
	\section{Simulation Results}\label{sec:4}
	We assess the system performance in terms of the DL achievable sum rate via numerical simulations. The achievable rate of some generic user $k$ is evaluated according to
	\begin{align}\label{eq:AchievRate}
		R_k=\tau \mathbb{E}\left[\log_2\left(1+\frac{|\bs{h}_k^\HH \bs{p}_k|^2}{\sum_{j\neq k}|\bs{h}_k^\HH \bs{p}_j|^2+\sigma^2}\right)\right]
	\end{align}
	where $N_\text{cov}=100$ channel covariance matrix realisations and $N_\text{ch}=200$ channel realisations are used to evaluate the expectation in \eqref{eq:AchievRate}. The pre-log factor $\tau=1-\frac{T_\tdl}{T_\text{coh}}$ accounts for dedicating $T_\tdl$ channel uses for DL training out of the $T_\text{coh}$ channel uses, which constitute the channel coherence interval. In our simulations, we assume the length of the channel coherence interval to be $T_\text{coh}=200$ symbols.
	
	We compare the rate results obtained using zero-forcing precoding to those when matched filters (or maximum ratio transmission) are applied. In the latter case, the precoding matrix is matched to the channel estimate, i.e., $\bs{P}^\text{MF}=\delta' \hat{\bs{H}}$, with the normalization scalar $\delta'=\frac{1}{\text{tr}( \hat{\bs{H}}^\HH  \hat{\bs{H}})}$.\\
	As a benchmark, we consider zero-forcing and matched filter precoding based on perfect channel knowledge (referred to by genie-aided). We recall, that in the asymptotic limit $P_\tdl\rightarrow\infty$, the zero-forcing precoder based on perfectly known channels is optimal, since the inter-user interference is completely removed.
	
	We consider a setup where the channels are generated using the QUAsi Deterministic RadIo channel GenerAtor (QuaDRiGa) \cite{quadriga}. The channel covariance matrices and means can be obtained by fitting a Gaussian mixture model (GMM) to the DL training observations \cite{GMM2}. The covariance matrix and the mean vector of the GMM component corresponding to the highest responsibility is used respectively as the channel covariance matrix $\bs{C}_k$ and channel mean vector $\bs{\mu}_k$ of the user $k$. Since the channels have a non-zero mean, the LMMSE channel estimate of user $k$ is given by
	\begin{equation*}
		\hat{\bs{h}}_k^\text{LMMSE}=\bs{G}_k(\bs{y}_k-\bs{\Phi}^\HH\bs{\mu}_k)+\bs{\mu}_{k}.
	\end{equation*} 
	The LS estimate does not depend on the channel statistics and its expression remains therefore unchanged [see \eqref{eq:LS}] even if the channel vector is non-zero mean.
	
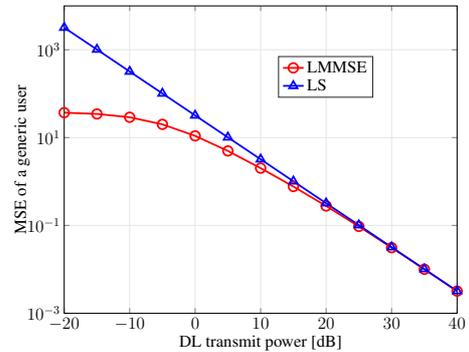
\begin{figure}[h!]
		\centering
		\scalebox{0.45}{
%
%
\begin{tikzpicture}

\begin{axis}[%
	width=4.521in,
height=3.566in,
at={(0.758in,0.481in)},
scale only axis,
xmin=-20,
xmax=40,
xlabel style={font=\color{white!15!black}, font=\Large},
xlabel={DL transmit power [dB]},
xticklabel style={font=\Large},
yticklabel style={font=\Large},
ymode=log,
ymin=0.001,
ymax=10000,
ylabel style={font=\color{white!15!black}, font=\Large},
ylabel={MSE of a generic user},
axis background/.style={fill=white},
title style={font=\bfseries},
xtick={-20,-10,0,10,20,30,40},
ytick={0.001,0.1,10,1000},
xmajorgrids,
ymajorgrids,
grid style={white!90!black},
legend style={at={(0.545,0.702)}, anchor=south west, legend cell align=left, align=left, draw=white!15!black, font=\Large}
]
\addplot [color=red, line width=1.5pt,mark=o, mark options={solid,red}, mark size=4pt]
   table[row sep=crcr]{%
 	-20	36.9866749720309\\
 	-15	34.5582996579316\\
 	-10	29.061679647342\\
 	-5	20.0251424021645\\
 	0	10.9489194122471\\
 	5	4.967191605122\\
 	10	2.01501644146196\\
 	15	0.768259975082707\\
 	20	0.276906301536424\\
 	25	0.0948714717419685\\
 	30	0.0311284038303899\\
 	35	0.0100297341985182\\
 	40	0.00318986593084055\\
};
\addlegendentry{LMMSE}

\addplot [color=blue, line width=1.5pt, mark=triangle, mark options={solid, blue}, mark size=4pt]
 table[row sep=crcr]{%
 	-20	3198.22083091381\\
 	-15	1013.41237929596\\
 	-10	319.566440570623\\
 	-5	101.213502137386\\
 	0	32.0042049989599\\
 	5	10.1277245771542\\
 	10	3.20806757973881\\
 	15	1.01206197941099\\
 	20	0.31994591876433\\
 	25	0.101414278037625\\
 	30	0.0319521295209535\\
 	35	0.0101217415101322\\
 	40	0.00319903869545601\\
};
\addlegendentry{LS}

\end{axis}

\end{tikzpicture}%
}
		\caption{Average \textbf{MSE} vs. DL transmit power $P_\tdl$ of the LMMSE channel estimate and the LS estimate for $M=32$~antennas, $T_\tdl= \textbf{32}$  \textbf{DL pilots} and $K=5$ users.}
		\label{fig:MSE32}
	\end{figure}

	In the following, we assume that the BS is equipped with $M=32$ antennas in a uniform linear array configuration and serves $K=5$ single-antenna users. Different numbers of DL training sequences are considered. First, the contamination-free case is considered, i.e., $T_\tdl=M=32$. In this case, the pilot matrix $\bs{\Phi}$ is square and both estimates, i.e., the LS and LMMSE estimates converge to the true channel for $P_\tdl\rightarrow\infty$. This can be seen in Fig.~\ref{fig:MSE32}, where the MSE $\mathbb{E}[\|\bs{h}_k-\hat{\bs{h}}_k\|_2^2]$ of a generic user $k$ is plotted. It can be observed that in the low power region, the LS channel estimate's quality is worse than LMMSE. The gap between the MSE of these estimates vanishes for high transmit power (starting from approx.~$20$~dB). For both estimators, the MSE converges to 0 due to the ability of perfect channel reconstruction when the training noise becomes negligible. 
	The results of the MSE are reflected in the rate performance as depicted in Fig.~\ref{fig:SR32}, where we plot the achievable sum rate with respect to the DL transmit power $P_\tdl$. We observe that the LMMSE estimate leads to a better performance in the low power regime. The gap to the LS estimate vanishes as the DL transmit power increases. Since both estimates converge to the true channel for $P_\tdl\rightarrow\infty$, the slope of the rate curves obtained for zero-forcing based on the LS and LMMSE estimates are approximately the same as in the genie-aided case.\\
	Furthermore, we observe that while the matched filter precoder leads to the best results at low transmit power, its performance is poor at medium to high transmit powers, since no interference mitigation takes place. 
	\begin{figure}[h!]
		\centering
		\scalebox{0.45}{
%
%
\begin{tikzpicture}

\begin{axis}[%
	width=4.521in,
height=3.566in,
at={(0.758in,0.481in)},
scale only axis,
xmin=-20,
xmax=40,
xlabel style={font=\color{white!15!black}, font=\Large},
xlabel={DL transmit power [dB]},
xticklabel style={font=\Large},
yticklabel style={font=\Large},
ymin=0,
ymax=70,
ylabel style={font=\color{white!15!black},font=\Large},
ylabel={Average achievable sum rate [bpcu]},
axis background/.style={fill=white},
title style={font=\bfseries},
xtick={-20,-10,0,10,20,30,40},
ytick={0,20,40,60},
grid style={white!90!black},
xmajorgrids,
ymajorgrids,
legend style={at={(0.245,0.602)}, anchor=south west, legend cell align=left, align=left, draw=white!15!black,font=\Large}
]
\addplot [color=green, line width=1.5pt, mark=square, mark options={solid, green}, mark size=4pt]
 table[row sep=crcr]{%
	-20	0.274165915880537\\
	-15	0.824039745854059\\
	-10	2.275426551020288\\
	-5	5.349958347035853\\
	0	10.240140714105960\\
	5	16.353899419496770\\
	10	23.025690754478497\\
	15	29.901723346197695\\
	20	36.845737899660980\\
	25	43.811615154576940\\
	30	50.784443379658626\\
	35	57.759473461124060\\
	40	64.735200208300780\\
};
\addlegendentry{ZF genie-aided}

\addplot [color=red, line width=1.5pt, mark=o, mark options={solid,red}, mark size=4pt]
 table[row sep=crcr]{%
 	-20	0.036533592973402\\
 	-15	0.146683606928279\\
 	-10	0.659715686094339\\
 	-5	2.636567716935289\\
 	0	7.228421558613150\\
 	5	13.571736774316475\\
 	10 20.301056673556175	\\
 	15	27.043013444856190\\
 	20	33.799103057471804\\
 	25	40.602590549595980\\
 	30	47.496831366886184\\
 	35	54.423222147586070\\
 	40	61.387029602311810\\
};
\addlegendentry{ZF LMMSE}

\addplot [color=blue, line width=1.5pt, mark=triangle, mark options={solid, blue}, mark size=4pt]
   table[row sep=crcr]{%
  	-20	0.019054834507344\\
  	-15	0.089752070565546\\
  	-10	0.484517044257832\\
  	-5	2.185122346442053\\
  	0	6.436118635144590\\
  	5	12.717617012103023\\
  	10	19.594044527648975\\
  	15	26.525040364305420\\
  	20	33.492469631811450\\
  	25	40.454076894911060\\
  	30	47.438414511418730\\
  	35	54.402851496707440\\
  	40 61.380685395025665\\
};
\addlegendentry{ZF LS}

\addplot [color=green, dashed, line width=1.5pt, mark=square, mark options={solid, green}, mark size=4pt]
  table[row sep=crcr]{%
 	-20	0.606587248757659\\
 	-15	1.63684289208529\\
 	-10	3.73743776934987\\
 	-5	6.8840885659152\\
 	0	10.276804426428\\
 	5	13.001665947741\\
 	10	14.7355503194044\\
 	15	15.6587811110804\\
 	20	16.0821965945213\\
 	25	16.2519947799315\\
 	30	16.3130266123474\\
 	35	16.3334607728113\\
 	40	16.340064161997\\
};
\addlegendentry{MF genie-aided}

\addplot [color=red, dashed, line width=1.5pt, mark=o, mark options={solid,red}, mark size=4pt]
  table[row sep=crcr]{%
	-20	0.073204727379062\\
	-15	0.310058511565637\\
	-10	1.258238571949683\\
	-5	3.673423673824808\\
	0	7.114074414162348\\
	5	10.117236693060420\\
	10	12.027549912701737\\
	15	13.019574581756903\\
	20	13.464687750655258\\
	25	13.637694972276313\\
	30	13.697660189208070\\
	35	13.718931292568628\\
	40	13.725132707960016\\
};
\addlegendentry{MF LMMSE}

\addplot [color=blue, dashed, line width=1.5pt, mark=triangle, mark options={solid, blue}, mark size=4pt]
    table[row sep=crcr]{%
  	-20	0.020261084349017\\
  	-15	0.101598223648290\\
  	-10	0.579762541455426\\
  	-5	2.524584905984509\\
  	0	6.245060905082613\\
  	5	9.720867013825043\\
  	10	11.868399261754224\\
  	15	12.951780794132207\\
  	20	13.437275421881220\\
  	25	13.627591191106871\\
  	30	13.694366218197253\\
  	35	13.717894742336064\\
  	40	13.724806602211656\\
};
\addlegendentry{MF LS}

\end{axis}

\end{tikzpicture}%
}
		\caption{Average achievable \textbf{sum rate} vs. DL transmit power $P_\tdl$ when zero-forcing (ZF) precoding and matched filter (MF) precoding are applied based on the true channel, the LMMSE channel estimate and the LS estimate for $M=32$~antennas, $T_\tdl=\textbf{32}$ \textbf{DL pilots} and $K=5$ users.}
		\label{fig:SR32}
	\end{figure}
	
\begin{figure}[h!]
		\centering
		\scalebox{0.45}{
%
%
\begin{tikzpicture}

\begin{axis}[%
	width=4.521in,
height=3.566in,
at={(0.758in,0.481in)},
scale only axis,
xmin=-20,
xmax=40,
xlabel style={font=\color{white!15!black}, font=\Large},
xlabel={DL transmit power [dB]},
xticklabel style={font=\Large},
yticklabel style={font=\Large},
ylabel style={font=\color{white!15!black}, font=\Large},
ymin=10,
ymax=1614.35460811052,
xtick={-20,-10,0,10,20,30,40},
ytick={10,100,1000,10000},
ymode=log,
ylabel={MSE of a generic user},
axis background/.style={fill=white},
title style={font=\bfseries},
xmajorgrids,
ymajorgrids,
grid style={white!90!black},
legend style={at={(0.545,0.702)}, anchor=south west, legend cell align=left, align=left, draw=white!15!black, font=\Large}
]
\addplot [color=red, line width=1.5pt,mark=o, mark options={solid,red}, mark size=4pt]
  table[row sep=crcr]{%
-20	34.8436889555221\\
-15	33.2415714634675\\
-10	29.815187241855\\
-5	24.4514460589039\\
0	19.1731030045421\\
5	15.6704183547886\\
10	13.9407719667845\\
15	13.2004133345804\\
20	12.9030116826158\\
25	12.7961151775308\\
30	12.7572698644682\\
35	12.7450252658772\\
40	12.7409192954607\\
};
\addlegendentry{LMMSE}

\addplot [color=blue, line width=1.5pt, mark=triangle, mark options={solid, blue}, mark size=4pt]
  table[row sep=crcr]{%
-20	1614.35460811052\\
-15	518.346514681775\\
-10	173.477765788526\\
-5	64.0178749254485\\
0	29.3838259910031\\
5	18.3987743833744\\
10	14.9353737852753\\
15	13.840469714469\\
20	13.4945301683701\\
25	13.3852570546787\\
30	13.350604815527\\
35	13.3396569119348\\
40	13.3361959605664\\
};
\addlegendentry{LS}

\end{axis}
\end{tikzpicture}%
}
		\caption{Average  \textbf{MSE} vs. DL transmit power $P_\tdl$ of the LMMSE channel estimate and the LS estimate for $M=32$~antennas, $T_\tdl= \textbf{16}$  \textbf{DL pilots} and $K=5$ users.}
		\label{fig:MSE16}
	\end{figure}
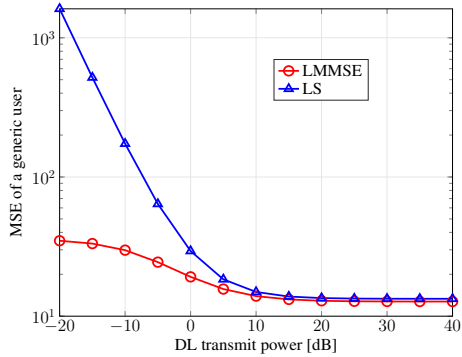

\begin{figure}[h!]
		\centering
		\scalebox{0.45}{
%
%
\begin{tikzpicture}

\begin{axis}[%
	width=4.521in,
height=3.566in,
at={(0.758in,0.481in)},
scale only axis,
xmin=-20,
xmax=40,
xlabel style={font=\color{white!15!black},font=\Large},
xlabel={DL transmit power [dB]},
xticklabel style={font=\Large},
yticklabel style={font=\Large},
ymin=0,
ymax=80,
ylabel style={font=\color{white!15!black},font=\Large},
xtick={-20,-10,0,10,20,30,40},
ytick={0,20,40,60,80},
ylabel={Average achievable sum rate [bpcu]},
axis background/.style={fill=white},
title style={font=\bfseries},
grid style={white!90!black},
xmajorgrids,
ymajorgrids,
legend style={at={(0.245,0.602)}, anchor=south west, legend cell align=left, align=left, draw=white!15!black,font=\Large}
]
\addplot [color=green, line width=1.5pt, mark=square, mark options={solid, green}, mark size=4pt]
  table[row sep=crcr]{%
 	-20	0.284310720254533\\
 	-15	0.856799396789546\\
 	-10	2.380309977727638\\
 	-5	5.651272001322409\\
 	0	10.926420680764425\\
 	5	17.580867488660890\\
 	10	24.872077120202760\\
 	15	32.397555289482625\\
 	20	40.001153477227660\\
 	25	47.629889886114680\\
 	30	55.266620218908976\\
 	35	62.905882939401074\\
 	40	70.545946921014480\\
};
\addlegendentry{ZF genie-aided}

\addplot [color=red, line width=1.5pt,mark=o, mark options={solid,red}, mark size=4pt]
 table[row sep=crcr]{%
	-20	0.029488496014736\\
	-15	0.092603295098160\\
	-10	0.306604382276183\\
	-5	0.997104273255517\\
	0	2.770553179759368\\
	5	5.961170259463128\\
	10	10.130937745331137\\
	15	14.394133445944945\\
	20	18.036973253767550\\
	25	20.643865873694676\\
	30	22.285605461114145\\
	35	23.208006747003230\\
	40	23.674069777039843\\
};
\addlegendentry{ZF LMMSE}

\addplot [color=blue, line width=1.5pt, mark=triangle, mark options={solid, blue}, mark size=4pt]
  table[row sep=crcr]{%
	-20	0.020833771122540\\
	-15	0.083287089432090\\
	-10	0.366505060674639\\
	-5	1.395111540238242\\
	0	3.776137013060959\\
	5	7.329299439722485\\
	10	11.550560608679424\\
	15	16.521611806201650\\
	20	22.497495439764805\\
	25	29.280525074780314\\
	30	36.588719645120440\\
	35	44.088827961139770\\
	40	51.698503028922870\\
};
\addlegendentry{ZF LS}

\addplot [color=green, dashed, line width=1.5pt, mark=square, mark options={solid, green}, mark size=4pt]
  table[row sep=crcr]{%
 	-20	0.547414084680977\\
 	-15	1.477181793021296\\
 	-10	3.379230876435082\\
 	-5	6.257284394721149\\
 	0	9.412188287150416\\
 	5	11.987491997662692\\
 	10	13.637108481046870\\
 	15	14.509967043906112\\
 	20	14.904987096857825\\
 	25	15.061575587532843\\
 	30	15.117523917795310\\
 	35	15.136217432908763\\
 	40	15.142254668864993\\
};
\addlegendentry{MF genie-aided}

\addplot [color=red, dashed, line width=1.5pt, mark=o, mark options={solid,red}, mark size=4pt]
   table[row sep=crcr]{%
 	-20	0.071895946447800\\
 	-15	0.284701596881282\\
 	-10	1.060743204461315\\
 	-5	2.933359401556905\\
 	0	5.567648135656230\\
 	5	7.848370317135776\\
 	10	9.292058205440268\\
 	15	10.083365229522718\\
 	20	10.502174274102657\\
 	25	10.708591754209692\\
 	30	10.801826291169416\\
 	35	10.838068337243874\\
 	40	10.851556050309380\\
};
\addlegendentry{MF LMMSE}

\addplot [color=blue, dashed, line width=1.5pt, mark=triangle, mark options={solid, blue}, mark size=4pt]
  table[row sep=crcr]{%
	-20	0.023157074124306\\
	-15	0.105110226926260\\
	-10	0.526939967568034\\
	-5	2.047302396292351\\
	0	4.791757884980964\\
	5	7.351228173050110\\
	10	8.909527385671074\\
	15	9.670245472357516\\
	20	10.005646247413077\\
	25	10.145078681546329\\
	30	10.199929914413772\\
	35	10.220576083513336\\
	40	10.228201994228916\\
};
\addlegendentry{MF LS}

\end{axis}

\end{tikzpicture}%
}
		\caption{Average achievable \textbf{sum rate} vs. DL transmit power $P_\tdl$ when zero-forcing (ZF) precoding and matched filter (MF) precoding are applied based on the true channel, the LMMSE channel estimate and the LS estimate for $M=32$~antennas, $T_\tdl=\textbf{16}$ \textbf{DL pilots} and $K=5$ users.}
		\label{fig:SR16}
	\end{figure}
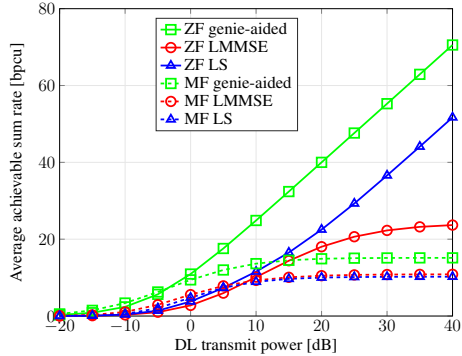

	
	We now consider the case where the number of DL pilots used for channel estimation is less than the number required for a contamination-free estimation, i.e., $T_\tdl<M$. In Fig.~\ref{fig:MSE16}, we plot the MSE of a generic user, when $T_\tdl=16$ and $M=32$. Similarly to the previous case, the LMMSE estimate yields a lower MSE than the LS estimate for the whole power range as expected. Interestingly, both estimates have slightly different error floors due to the pilot contamination effect. Additionally, there exists a saturation of the MSEs in both cases which is caused by the systematic error that does not vanish even when increasing the DL transmit power.
	Intuitively, one would expect that basing the zero-forcing precoder on the channel estimate with better quality would lead to a better performance in terms of the system throughput. However, this is not the case as can be seen in Fig~\ref{fig:SR16}, where the achievable sum rate is plotted. Contrarily to the contamination-free case, we observe that the rate curve obtained with zero-forcing based on the LMMSE channel estimate saturates due to the residual interference that we discussed in the previous section. The simulation results also confirm our analytical results concerning the interference-free transmission for zero-forcing based on the LS channel estimate. This can be seen by the non-saturating rate curve that exhibits the same slope as the perfect CSI case.
	
\begin{figure}[h!]
		\centering
		\scalebox{0.45}{\begin{tikzpicture}
	
	\begin{axis}[%
		width=4.521in,
		height=3.566in,
		at={(0.758in,0.481in)},
		scale only axis,
		xmin=-20,
		xmax=40,
		xlabel style={font=\color{white!15!black},font=\Large},
		xlabel={DL transmit power [dB]},
		xticklabel style={font=\Large},
		yticklabel style={font=\Large},
		ymin=0,
		ymax=70,
		ylabel style={font=\color{white!15!black},font=\Large},
		ylabel={Average achievable sum rate [bpcu]},
		axis background/.style={fill=white},
		xtick={-20,-10,0,10,20,30,40},
		ytick={0,20,40,60},
		title style={font=\bfseries},
		xmajorgrids,
		ymajorgrids,
		grid style={white!90!black},
legend style={at={(0.245,0.602)}, anchor=south west, legend cell align=left, align=left, draw=white!15!black, font=\Large}
		]
		
		\addplot [color=red, line width=1.5pt,mark=o,mark options={solid,red}, mark size=4pt]
		 table[row sep=crcr]{%
			-20	0.036533592973402\\
			-15	0.146683606928279\\
			-10	0.659715686094339\\
			-5	2.636567716935289\\
			0	7.228421558613150\\
			5	13.571736774316475\\
			10 20.301056673556175	\\
			15	27.043013444856190\\
			20	33.799103057471804\\
			25	40.602590549595980\\
			30	47.496831366886184\\
			35	54.423222147586070\\
			40	61.387029602311810\\
		};
		\addlegendentry{$T_\text{dl}=32$}

		\addplot [color=red!75!white, dashed, line width=1.5pt,mark=o,mark options={solid,red!75!white}, mark size=4pt]
	  table[row sep=crcr]{%
		-20	0.035954308087111\\
		-15	0.141710946937311\\
		-10	0.625315981464578\\
		-5	2.504465965731189\\
		0	6.972163071670355\\
		5	13.245460128696477\\
		10	19.895936070738205\\
		15	26.387523406249734\\
		20	32.512968022612290\\
		25	38.067491049472140\\
		30	42.847314942532954\\
		35	46.733202034228730\\
		40	49.751518022448470\\
		};
		\addlegendentry{$T_\text{dl}=31$}
		
		\addplot [color=red!60!white, loosely dashdotted, line width=1.5pt,mark=o,mark options={solid,red!60!white}, mark size=4pt]
 table[row sep=crcr]{%
	-20	0.029488496014736\\
	-15	0.092603295098160\\
	-10	0.306604382276183\\
	-5	0.997104273255517\\
	0	2.770553179759368\\
	5	5.961170259463128\\
	10	10.130937745331137\\
	15	14.394133445944945\\
	20	18.036973253767550\\
	25	20.643865873694676\\
	30	22.285605461114145\\
	35	23.208006747003230\\
	40	23.674069777039843\\
		};
		\addlegendentry{$T_\text{dl}=16$}
		
		\addplot [color=red!45!white, dotted, line width=1.5pt,mark=o,mark options={solid,red!45!white}, mark size=4pt]
	  table[row sep=crcr]{%
		-20	0.027042915211290\\
		-15	0.079649164857911\\
		-10	0.231190122760271\\
		-5	0.653213107898740\\
		0	1.700102793763344\\
		5	3.797281168213267\\
		10	6.822538466133485\\
		15	9.797076006199967\\
		20	11.960742538300416\\
		25	13.220329957357151\\
		30	13.860787663124254\\
		35	14.139742236933024\\
		40	14.243890081425695\\
		};
		\addlegendentry{$T_\text{dl}=8$}
		
		\addplot [color=red!30!white, loosely dotted, line width=1.5pt,mark=o,mark options={solid,red!30!white}, mark size=4pt]
	  table[row sep=crcr]{%
		-20	0.028673380226304\\
		-15	0.084891577817070\\
		-10	0.233730771921804\\
		-5	0.585270407417175\\
		0	1.221950391351455\\
		5	2.128551734475944\\
		10	3.114244763892071\\
		15	3.913395728164662\\
		20	4.426518084910005\\
		25	4.699528931470919\\
		30	4.818322349278632\\
		35	4.859819699921175\\
		40	4.871789844991267\\
		};
		\addlegendentry{$T_\text{dl}=3$}

	\end{axis}

\end{tikzpicture}%
		
		}
		\caption{Average achievable \textbf{sum rate} vs. DL transmit power $P_\tdl$ when zero-forcing (ZF) precoding is applied based on the \textbf{LMMSE} channel estimate for a system with $M=32$~antennas, $K=5$ users and \textbf{different numbers of DL pilots $T_\tdl$}.}
		\label{fig:ZF_MMSE}
	\end{figure}
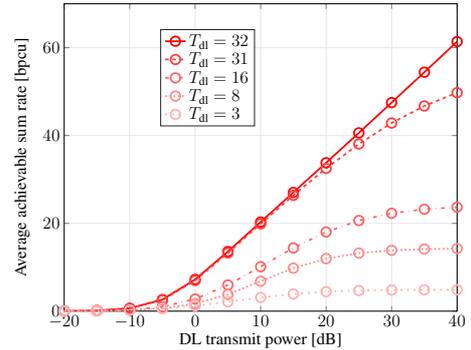

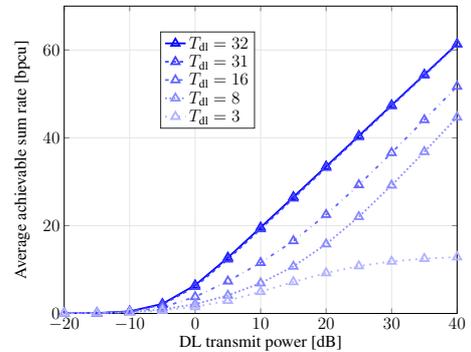
\begin{figure}[h!]
		\centering
		\scalebox{0.45}{\begin{tikzpicture}
	
	\begin{axis}[%
		width=4.521in,
		height=3.566in,
		at={(0.758in,0.481in)},
		scale only axis,
		xmin=-20,
		xmax=40,
		xlabel style={font=\color{white!15!black},font=\Large},
		xlabel={DL transmit power [dB]},
		xticklabel style={font=\Large},
		yticklabel style={font=\Large},
		ymin=0,
		ymax=70,
		xtick={-20,-10,0,10,20,30,40},
		ytick={0,20,40,60},
		ylabel style={font=\color{white!15!black},font=\Large},
		ylabel={Average achievable sum rate [bpcu]},
		axis background/.style={fill=white},
		title style={font=\bfseries},
		xmajorgrids,
		ymajorgrids,
		grid style={white!90!black},
legend style={at={(0.245,0.602)}, anchor=south west, legend cell align=left, align=left, draw=white!15!black,font=\Large}
		]

\addplot [color=blue, line width=1.5pt,mark=triangle, mark options={solid, blue}, mark size=4pt]
 table[row sep=crcr]{%
	-20	0.019054834507344\\
-15	0.089752070565546\\
-10	0.484517044257832\\
-5	2.185122346442053\\
0	6.436118635144590\\
5	12.717617012103023\\
10	19.594044527648975\\
15	26.525040364305420\\
20	33.492469631811450\\
25	40.454076894911060\\
30	47.438414511418730\\
35	54.402851496707440\\
40 61.380685395025665\\
};
\addlegendentry{$T_\text{dl}=32$}

\addplot [color=blue!75!white, dashed, line width=1.5pt,mark=triangle, mark options={solid, blue!75!white}, mark size=4pt]
table[row sep=crcr]{%
	-20	0.017931152782924\\
	-15	0.082781988125403\\
	-10	0.441361257271640\\
	-5	2.017560983887663\\
	0	6.105252670537186\\
	5	12.334447666502410\\
	10	19.238831835881346\\
	15	26.221282500677557\\
	20	33.212939408494400\\
	25	40.213163480643500\\
	30	47.240053173188430\\
	35	54.249450011093586\\
	40	61.261880928490854\\
};
\addlegendentry{$T_\text{dl}=31$}

\addplot [color=blue!60!white, loosely dashdotted, line width=1.5pt,mark=triangle, mark options={solid, blue!60!white}, mark size=4pt]
 table[row sep=crcr]{%
	-20	0.020833771122540\\
	-15	0.083287089432090\\
	-10	0.366505060674639\\
	-5	1.395111540238242\\
	0	3.776137013060959\\
	5	7.329299439722485\\
	10	11.550560608679424\\
	15	16.521611806201650\\
	20	22.497495439764805\\
	25	29.280525074780314\\
	30	36.588719645120440\\
	35	44.088827961139770\\
	40	51.698503028922870\\
};
\addlegendentry{$T_\text{dl}=16$}

\addplot [color=blue!45!white, dotted, line width=1.5pt,,mark=triangle, mark options={solid, blue!45!white}, mark size=4pt]
 table[row sep=crcr]{%
	-20	0.028657187955332\\
	-15 0.093948517789497\\
	-10	0.305909869208019\\
	-5	0.896944224231537\\
	0	2.122373416480550\\
	5	4.108267156992874\\
	10	6.907180081041965\\
	15	10.708648454132064\\
	20	15.792131595346560\\
	25	22.023596822887487\\
	30	29.184420312098590\\
	35	36.817021181016480\\
	40	44.686631596015390\\
};
\addlegendentry{$T_\text{dl}=8$}

\addplot [color=blue!30!white,loosely dotted, line width=1.5pt,mark=triangle, mark options={solid, blue!30!white}, mark size=4pt]
  table[row sep=crcr]{%
	-20	0.029009706552033\\
	-15	0.087564823136111\\
	-10	0.248976781535496\\
	-5	0.652326963193193\\
	0	1.492109885363913\\
	5	2.924435858327987\\
	10	4.943233195976834\\
	15 7.179462150267304\\
	20	9.215060266008388\\
	25	10.796312056486900\\
	30	11.849716784229141\\
	35	12.480111098196277\\
	40 12.830051913150903\\
};
\addlegendentry{$T_\text{dl}=3$}

\end{axis}

\end{tikzpicture}%
}
		\caption{Average achievable \textbf{sum rate} vs. DL transmit power $P_\tdl$ when zero-forcing (ZF) precoding is applied based on the \textbf{LS} channel estimate for a system with $M=32$~antennas, $K=5$ users and \textbf{different numbers of DL pilots $T_\tdl$}.}
		\label{fig:ZF_LS}
	\end{figure}

%
		In Figs.~\refeq{fig:ZF_MMSE} and \ref{fig:ZF_LS}, we plot the achievable sum rate for zero-forcing precoding based on the LMMSE and LS channel estimates, respectively, for different number of DL training pilots. On the one hand, it can be observed that the approach based on the LS channel estimate eliminates the inter-user interference in the high power regime with a degradation of the performance as the number of DL pilots decreases. However, the rate curve only saturates when $T_\tdl=3<K$. In this case, the zero-forcing precoder based on the LS estimates fails to mitigate the inter-user interference. On the other hand, zero-forcing based on the LMMSE estimates exhibits a saturation of the rates for all $T_\tdl<M$ at high transmit power values, which is due to the residual interference.
	
\begin{figure}[h!]
		\centering
		\scalebox{0.45}{\begin{tikzpicture}
	
	\begin{axis}[%
		width=4.521in,
    	height=3.566in,
	    at={(0.758in,0.481in)},
		scale only axis,
		xmin=-20,
		xmax=40,
		xlabel style={font=\color{white!15!black},font=\Large},
		xlabel={DL transmit power [dB]},
		xticklabel style={font=\Large},
		yticklabel style={font=\Large},
		ymin=0,
		ymax=70,
		ylabel style={font=\color{white!15!black},font=\Large},
		ylabel={Average achievable sum rate [bpcu]},
		axis background/.style={fill=white},
		title style={font=\bfseries},
		xtick={-20,-10,0,10,20,30},
		ytick={0,10,20,30,40,50,60},
		grid style={white!90!black},
		xmajorgrids,
		ymajorgrids,
legend style={at={(0.145,0.702)}, anchor=south west, legend cell align=left, align=left, draw=white!15!black,font=\Large}
		]

\addplot [color=red, line width=1.5pt,mark=o, mark options={solid, red}, mark size=4pt]
  table[row sep=crcr]{%
	-20	0.0354023393593034\\
	-15	0.140564620427981\\
	-10	0.623688755661904\\
	-5	2.52264867135903\\
	0	7.03875775267946\\
	5	13.351547839732\\
	10	20.0896884325781\\
	15	26.820932235683\\
	20	33.5516255130638\\
	25	40.3452436642084\\
	30	47.217715676727\\
	35	54.1497435575379\\
	40	61.107402771109\\
};
\addlegendentry{ZF LMMSE}

\addplot [color=blue, line width=1.5pt,mark=triangle, mark options={solid, blue}, mark size=4pt]
 table[row sep=crcr]{%
	-20	0.0185155287259004\\
	-15	0.0864233285832097\\
	-10	0.45712867241582\\
	-5	2.08768230653472\\
	0	6.22655408604147\\
	5	12.4459320678784\\
	10	19.3198417644867\\
	15	26.2521241055996\\
	20	33.2090744718066\\
	25	40.1752017187707\\
	30	47.1506333703984\\
	35	54.1255681820938\\
	40	61.0985171537606\\
};
\addlegendentry{ZF LS}

\addplot [color=red, line width=1.5pt, mark=o, mark options={solid, red}, mark size=4pt]
 table[row sep=crcr]{%
	-20	0.0286158882547268\\
	-15	0.0871787786029422\\
	-10	0.272877729485321\\
	-5	0.857688827878059\\
	0	2.35077085046492\\
	5	5.19197956720102\\
	10	9.07391576416601\\
	15	13.0792083202714\\
	20	16.4486125533284\\
	25	18.8596912239932\\
	30	20.3078552311475\\
	35	21.0741654375868\\
	40	21.4531483673668\\
};

\addplot [color=blue, line width=1.5pt,mark=triangle, mark options={solid, blue}, mark size=4pt]
  table[row sep=crcr]{%
 	-20	0.0211888185097815\\
 	-15	0.0837381092772583\\
 	-10	0.359314876276818\\
 	-5	1.35996655439861\\
 	0	3.64653193870041\\
 	5	6.99771762485533\\
 	10	10.9190484203023\\
 	15	15.6049638177255\\
 	20	21.3394669055856\\
 	25	28.0122309690018\\
 	30	35.2550224614665\\
 	35	42.7492370230129\\
 	40	50.3237994995422\\
};
\end{axis}

\draw [black, thick] (10.5,6.4) ellipse [x radius=0.3, y radius=0.5];
\draw [black, thick] (10.5,4.2) ellipse [x radius=0.3, y radius=1];
\draw [-stealth, thick](9.5,7.3) -- (10.2,6.6);
\node at (9.5,7.5) {\Large $T_\text{dl}=32$};
\draw [-stealth, thick](11.5,2.6) -- (10.7,3.3);
\node at (12.5,2.5) {\Large $T_\text{dl}=16$};

\end{tikzpicture}}
		\caption{Average achievable \textbf{sum rate} vs. DL transmit power $P_\tdl$ when zero-forcing (ZF) precoding is applied based on the true channel, the LMMSE channel estimate and the LS estimate for $M=32$~antennas and $K=5$ users, noisy feedback channel.}
		\label{fig:SR32FB}
	\end{figure}
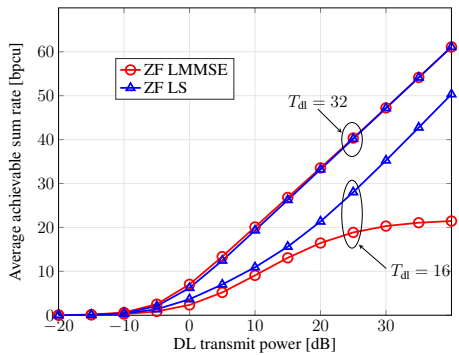

	We shall note that the asymptotic behavior of the zero-forcing precoder was based on the assumption of noiseless feedback. To model the noisy feedback, we consider an additional noise portion in $\bs{n}_k$ [cf.~\eqref{eq:yk}] that depends on the fixed UL power $P_\text{fb}$ that the users employ to send their channel observations through the feedback channel.  To this end, we evaluate the system performance for $P_\text{fb}=10~\text{dB}$ and we plot in Fig.~\ref{fig:SR32FB} the average achievable sum rate with respect to the DL transmit power. We see that there is a slight degradation of the performance due to the additional noise compared to Figs.~\ref{fig:SR32} and \ref{fig:SR16}, where a noiseless feedback channel was considered. However, the asymptotic behavior of the zero-forcing precoder based on the LS and LMMSE channel estimates remains the same. Thus, our results are not restricted to the noiseless feedback channel case.

\begin{figure}
	\centering
	\scalebox{0.45}{
%
%
\begin{tikzpicture}

\begin{axis}[%
	width=4.521in,
	height=3.566in,
	at={(0.758in,0.481in)},
	scale only axis,
	xmin=-20,
	xmax=40,
	xlabel style={font=\color{white!15!black}, font=\Large},
	xlabel={DL transmit power [dB]},
	xticklabel style={font=\Large},
	yticklabel style={font=\Large},
	ymin=0,
	ymax=120,
	ylabel style={font=\color{white!15!black},font=\Large},
	ylabel={Average achievable sum rate [bpcu]},
	axis background/.style={fill=white},
	title style={font=\bfseries},
	xtick={-20,-10,0,10,20,30,40},
	ytick={0,40,80,120},
	grid style={white!90!black},
	xmajorgrids,
	ymajorgrids,
	legend style={at={(0.245,0.602)}, anchor=south west, legend cell align=left, align=left, draw=white!15!black,font=\Large}
	]
\addplot [color=green, line width=1.5pt, mark=square, mark options={solid, green}, mark size=4pt]
  table[row sep=crcr]{%
-20	0.219625669953624\\
-15	0.680416930405528\\
-10	2.02598543656756\\
-5	5.47627910614766\\
0	12.4363300403066\\
5	22.9171266061113\\
10	35.5162502741806\\
15	49.0054197049087\\
20	62.807177002465\\
25	76.7113220189485\\
30	90.6482143114827\\
35	104.59549966227\\
40	118.546075349278\\
};
\addlegendentry{ZF genie-aided}

\addplot [color=red, line width=1.5pt, mark=o, mark options={solid,red}, mark size=4pt]
  table[row sep=crcr]{%
-20	0.00508663229976817\\
-15	0.0177494393711505\\
-10	0.0723515046602801\\
-5	0.353279419444662\\
0	1.77179259839909\\
5	6.60418077472595\\
10	15.4211296449853\\
15	24.6723115154924\\
20	31.3412590787857\\
25	34.8119164318772\\
30	36.055743643263\\
35	36.2302269955631\\
40	36.0579230106657\\
};
\addlegendentry{ZF LMMSE}

\addplot [color=blue, line width=1.5pt, mark=triangle, mark options={solid, blue}, mark size=4pt]
  table[row sep=crcr]{%
-20	0.0113550400708291\\
-15	0.0498084228030243\\
-10	0.24129378156827\\
-5	1.01985879323176\\
0	3.29515200960669\\
5	8.50954305216153\\
10	17.6378781748169\\
15	29.6172602259649\\
20	42.8385086600053\\
25	56.5473969394698\\
30	70.4174974324193\\
35	84.3438705752327\\
40	98.2833426098675\\
};
\addlegendentry{ZF LS}

\addplot [color=green, dashed, line width=1.5pt, mark=square, mark options={solid, green}, mark size=4pt]
  table[row sep=crcr]{%
-20	0.919965360666928\\
-15	2.25148034790142\\
-10	4.6391168244769\\
-5	8.00159025003306\\
0	11.6950988929489\\
5	14.787326737052\\
10	16.7072936629711\\
15	17.6042915630775\\
20	17.9449942966686\\
25	18.060730337139\\
30	18.0982488807479\\
35	18.110209984696\\
40	18.1140022507134\\
};
\addlegendentry{MF genie-aided}

\addplot [color=red, dashed, line width=1.5pt, mark=o, mark options={solid,red}, mark size=4pt]
  table[row sep=crcr]{%
-20	0.101450685125111\\
-15	0.38268632951059\\
-10	1.27523830658529\\
-5	3.12264814357105\\
0	5.67925709788039\\
5	8.44212919276417\\
10	10.7525703254344\\
15	12.1592631748691\\
20	12.8130223590024\\
25	13.0854757264613\\
30	13.199733005093\\
35	13.2556900632262\\
40	13.2916174639067\\
};
\addlegendentry{MF LMMSE}

\addplot [color=blue, dashed, line width=1.5pt, mark=triangle, mark options={solid, blue}, mark size=4pt]
  table[row sep=crcr]{%
-20	0.0129791557033374\\
-15	0.0663257519709363\\
-10	0.380224092152124\\
-5	1.737360869808\\
0	4.83026064536366\\
5	8.52064385067967\\
10	11.0773529810605\\
15	12.2883780998923\\
20	12.7480399418798\\
25	12.905316776809\\
30	12.9569627670183\\
35	12.973286263735\\
40	12.9784497508373\\
};
\addlegendentry{MF LS}

\end{axis}

\end{tikzpicture}%
}
	\caption{Average achievable \textbf{sum rate} vs. DL transmit power $P_\tdl$ when zero-forcing (ZF) precoding and matched filter (MF) precoding are applied based on the true channel, the LMMSE channel estimate and the LS estimate for $M=64$~antennas, $T_\tdl=32$ and $K=10$ users, for the \textbf{urban micro} scenario.}
	\label{fig:SR32mu}
\end{figure}
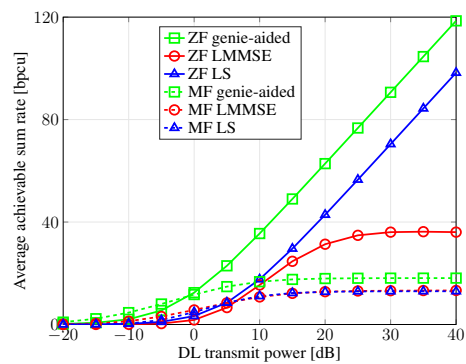

Finally, and in order to show that our results hold independently of the underlying scenario, we evaluate the system performance for the urban micro scenario from the 3GPP model \cite{ETSI}. We assume that the BS is equipped with $M=64$ antennas and is located at the center of a cell whose radius is given by $250$~m. 10 users are served by the BS and are placed uniformly at random inside the cell. The users' channels are non-line-of-sight (NLOS) and sampled from the Gaussian distribution with zero-mean and covariance matrix generated according to the multi-path channel model. Assuming $N_\text{path}=6$ main clusters and $N_\text{rays}=20$ rays in each cluster, the covariance matrix of user $k$ at carrier frequency $f$ is given by $
	\boldsymbol{C}_k=\eta_k \sum_{n=1}^{N_\text{paths}} \frac{\zeta_n}{N_\text{rays}}\sum_{m=1}^{N_\text{rays}} \boldsymbol{a}(\theta_{k,n,m},f)\boldsymbol{a}^\HH(\theta_{k,n,m},f)$,
where $\eta_k$ depends on the path-loss and $\zeta_n$ denotes the power of the $n$th cluster. $\boldsymbol{a}(\theta_{k,n,m},f)$ is the array steering vector and the $\theta_{k,n,m}$ is is the angle-of-arrival (AOA) of the $m$th ray inside the $n$th cluster for the $k$th user.\\
In Fig.~\ref{fig:SR32mu}, the average achievable rate results are plotted for the case $T_\tdl=M/2=32$. Obviously, the same observations made for the previous scenario hold here. While the zero-forcing precoder based on the LS channel estimate provides asymptotically full degrees of freedom, its counterpart based on the LMMSE estimate fails to remove the inter-user interference and the rate curve saturates at high transmit powers as observed in Fig.~\ref{fig:SR16}. Furthermore, we observe that the matched filter based on the LMMSE estimate outperforms the matched filter based on the LS estimate due to the better estimation quality especially in the low power regime. The matched filter performance in both cases is poor in the high power limit.

	\section{Conclusion}
	We have shown through analytical derivations and numerical simulations that zero-forcing precoding based on the LS channel estimate leads to an interference-free transmission in the asymptotic limit of high transmit powers, if the number of DL pilots is larger than or equal to the number of users. Contrarily, the zero-forcing precoder based on the better quality estimator, i.e., the LMMSE estimator, fails to eliminate the inter-user interference even at high transmit powers. Hence, we can conclude that a good quality estimator in terms of its MSE does not necessarily lead to a good performance in terms of the achievable system throughput. Besides being less computationally expensive than the LMMSE channel estimate, the LS channel estimate does not require the knowledge of the channel statistics that usually have to be additionally estimated at the transmitter.


\begin{thebibliography}{10}
		
		\bibitem{marzetta}
		T.~L. Marzetta.
		\newblock {Noncooperative Cellular Wireless with Unlimited Numbers of Base
			Station Antennas}.
		\newblock {\em IEEE Transactions on Wireless Communications}, 9(11):3590--3600,
		November 2010.
		
		\bibitem{marzetta2016fundamentals}
		T.~L. Marzetta, E.G. Larsson, H.~Yang, and H.Q. Ngo.
		\newblock {\em {Fundamentals of Massive {MIMO}}}.
		\newblock Cambridge University Press, 2016.
		
		\bibitem{mimo}
		E.~Biglieri, R.~Calderbank, A.~Constantinides, A.~Goldsmith, A.~Paulraj, and
		H.V. Poor.
		\newblock {\em {MIMO wireless communications}}.
		\newblock Cambridge university press, 2007.
		
		\bibitem{Joham}
		M.~Joham, W.~Utschick, and J.A. Nossek.
		\newblock {Linear Transmit Processing in MIMO Communications Systems}.
		\newblock {\em IEEE Transactions on Signal Processing}, 53(8):2700--2712, 2005.
		
		\bibitem{Stojnic}
		M.~Stojnic, H.~Vikalo, and B.~Hassibi.
		\newblock {Rate Maximization in Multi-Antenna Broadcast Channels with Linear
			Preprocessing}.
		\newblock {\em IEEE Transactions on Wireless Communications}, 5(9):2338--2342,
		2006.
		
		\bibitem{CaireShamai}
		G.~Caire and S.~Shamai.
		\newblock {On the Achievable Throughput of a Multiantenna Gaussian Broadcast
			Channel}.
		\newblock {\em IEEE Transactions on Information Theory}, 49(7):1691--1706,
		2003.
		
		\bibitem{Wiesel}
		A.~Wiesel, C.~Y. Eldar, and S.~Shamai.
		\newblock {Zero-Forcing Precoding and Generalized Inverses}.
		\newblock {\em IEEE Transactions on Signal Processing}, 56(9):4409--4418, 2008.
		
		\bibitem{Tse}
		D.~Tse and P.~Viswanath.
		\newblock {\em {Fundamentals of Wireless Communication}}.
		\newblock Cambridge University Press, USA, 2005.
		
		\bibitem{Donia_WSA20}
		D.~Ben Amor, M.~Joham, and W.~Utschick.
		\newblock {Bilinear Precoding for {FDD} Massive {MIMO} System with Imperfect
			Covariance Matrices}.
		\newblock In {\em {WSA 2020; 24th International ITG Workshop on Smart
				Antennas}}, pages 1--6, 2020.
		
		\bibitem{Extrapolation2}
		M.~{Barzegar Khalilsarai}, S.~{Haghighatshoar}, X.~{Yi}, and G.~{Caire}.
		\newblock {{FDD} Massive MIMO via UL/DL Channel Covariance Extrapolation and
			Active Channel Sparsification}.
		\newblock {\em IEEE Transactions on Wireless Communications}, 18(1):121--135,
		Jan 2019.
		
		\bibitem{quadriga}
		S.~Jaeckel, L.~Raschkowski, K.~Börner, and L.~Thiele.
		\newblock {QuaDRiGa: A 3-D Multi-Cell Channel Model With Time Evolution for
			Enabling Virtual Field Trials}.
		\newblock {\em IEEE Transactions on Antennas and Propagation},
		62(6):3242--3256, 2014.
		
		\bibitem{GMM2}
		M.~Koller, B.~Fesl, N.~Turan, and W.~Utschick.
		\newblock {An Asymptotically MSE-Optimal Estimator Based on Gaussian Mixture
			Models}.
		\newblock {\em IEEE Transactions on Signal Processing}, 70:4109--4123, 2022.
		
		\bibitem{ETSI}
		{Universal Mobile Telecommunications System (UMTS); Spatial Channel Model for
			Multiple Input Multiple Output {(MIMO)} Simulations (3GPP TR 25.996 version
			16.0.0 Release 16) }.
		\newblock Technical Report TR 125 996, ETSI 3rd Generation Partnership Project
		(3GPP).
		
	\end{thebibliography}
\end{document}